\documentclass[prb,twocolumn,superscriptaddress,a4paper]{revtex4}

\usepackage{graphicx}
\usepackage{amssymb}

\begin{document}

\title{Electronic and magnetic properties of the (001) surface of hole-doped manganites}

\author{H. Zenia}
\affiliation{Department of Physics and Astronomy, University of Sheffield, Sheffield, S3 7RH, UK }
\author{G. Banach}
\affiliation{ Daresbury Laboratory, Daresbury, Warrington WA4 4AD, UK }
\author{W.M. Temmerman}
\affiliation{ Daresbury Laboratory, Daresbury, Warrington WA4 4AD, UK }
\author{G.A. Gehring}
\affiliation{Department of Physics and Astronomy, University of Sheffield, Sheffield, S3 7RH, UK }

\begin{abstract}
The electronic and magnetic properties of ferromagnetic doped manganites are investigated by means of model
tight-binding and \textit{ab initio} self-interaction corrected local spin density approximation calculations.
It is found that the surface alone by breaking the cubic symmetry induces a difference in the occupation of 
the two $e_{g}$ orbitals at the surface. With \textit{ab initio} calculations we found  
surface localisation of one orbital and hence a change in the Mn valency from four in the bulk to three at the 
sub-surface. Different surface or disordered interface induced localisation of the orbitals are considered too with
respect to the nature and the strength of the magnetic exchange coupling between the surface/interface and the
bulk-like region.

\end{abstract}




\maketitle



\section{Introduction}

The half-metallic properties of La$_{1-x}$Sr$_{x}$MnO$_{3}$ (x$\sim$0.3) (LSMO) 
are of great importance for applications in spintronics. The 
tunnel magnetoresistance (TMR) of LSMO/SrTiO$_3$(STO)/LSMO junction
shows a magnetoresistance ratio in excess of 1800\% \cite{bowen}
and was, by these authors, attributed to half-metallicity.
While the magnetic impurities which might diffuse to the insulating layer could play an important role
in the tunnelling process \cite{guinea98} through the spin flip effect \cite{lyu98}, antiferromagnons at the 
interface \cite{guinea98} due to an antiferromagnetic interlayer coupling with the subsurface (or bulk) are found 
also to affect the MR in manganites tunnel junctions. The magnetic properties of these materials are highly 
sensitive to local crystal properties. The extrinsic strain field induced by lattice mismatch with the substrates or tunnel
barriers can be sufficient to severely degrade the ferromagnetic order in the surface layers which are critical for tunneling 
\cite{Jo00,Jo99}. 
Although other defects such as segregation of a particular species, like Sr in LSMO, at the interface alters the desired
electronic and magnetic properties we have not addressed the issue in this work. Good TMR is expected if the material is fully
polarised and half-metallic. Thus the occurrence of an antiferromagnetic layer or a localised layer at the surface will be very 
detrimental to tunneling. There is then a strong case for demanding to have both half-metallicity and 
ferromagnetic exchange at the 
interfaces between manganites and insulating barriers. Understanding the spin 
polarisation at the surface is then of major importance assuming that growing techniques could fabricate sharp well-defined 
interfaces \cite{bowen} and low diffusion rates of the magnetic ions into the insulating layer. In this paper we investigate
the conditions required for surface antiferromagnetism and/or localised surface states so as to make clear the physical conditions
for avoiding them.

On the basis of local spin density (LSD) band theory calculations, 
the origin of half-metallic character of manganese perovskites 
was discussed in several papers.
\cite{picket97,picket98,picketJAP,livesay,nadgorny,mazin}
These LSD calculations failed to obtain  
a half-metallic state and subsequently the possibility
of transport half-metallicity was raised by
Nadgorny{\it{et al.}}\cite{nadgorny} and Mazin.\cite{mazin}
It could equally well be argued that the
fascinating electronic and magnetic properties of LSMO, 
including colossal magnetoresistance (CMR), might indicate that 
the electronic structure is more complex than the standard 
band theory picture (see reviews \cite{coeyrev,tokura})
and might necessitate a better treatment of correlation effects.
Of particular importance would be to see if these correlation effects confirmed the half-metallicity
of these materials.

Recently two of us~\cite{banach} described how upon Sr doping of LaMnO$_3$ (LMO) the Mn valence increases from $3+$ to $4+$ 
by delocalizing the e$_g$ electron. These results therefore suggested that, in LSMO, Sr hole doping favours band formation 
instead of localisation. With this Sr doping no half-metallic state was obtained in LSMO. Rather, the calculations suggested, 
half-metallicity  is the consequence of remaining local Jahn-Teller distortions from the LMO parent material. This did go hand 
in hand with a mixed valence Mn$^{3+}$/Mn$^{4+}$ ground state which was discovered for Sr concentrations less than 
20\%.\cite{banach}

The importance of the local distortions in LSMO does suggest
that the surface properties of LSMO might be different from the bulk properties.
This could possibly have important consequences for the magnetoelectronic 
transport through an LSMO/STO interface. 
Surfaces of manganese perovskites were studied before:
Fillipetti and Pickett used a pseudopotential method to study the magnetic properties of  
the surface of CaMnO$_3$ (CMO) \cite{fillipetti1} 
and La$_{1-x}$Ca$_x$MnO$_3$ (LCMO) \cite{fillipetti2}
in the (001) direction, 
Evarestov {\it{et al.}} \cite{evarestov} used the Hartree-Fock approach 
to study the  surface (110) in LaMnO$_3$.
In particular the work of Fillipetti and Pickett~\cite{fillipetti1,fillipetti2}
stressed the importance of spin flip processes at the surface for the transport properties.

In this paper we deal with finite slabs of R$_{1-x}$A$_{x}$MnO$_{3}$ (where R and A are trivalent and divalent
ions respectively) to study the surface in a tight-binding model and Self-Interaction Corrected \cite{sic,rik} (SIC) LSDA 
calculations. The first allows us to study larger systems and more complicated magnetic configuration using a few parameters 
whereas the latter is parameter free and therefore more accurate but limited by the number of atoms that can be simulated. 

\section{Tight-binding methodology}

The active orbitals in a model calculation on manganites are the two degenerate $e_{g}$ orbitals separated by a 
``strong'' ligand field from the three low-lying $t_{2g}$ states. As we are concerned with a region of the phase 
diagram where most manganites are found to be in the FM metallic phase we use the Kondo-lattice type
model Hamiltonian \cite{Millis,calderon} using the two $e_{g}$ orbitals:

\begin{eqnarray}
H &=& - \frac{1}{2} \sum_{{\bf ia} \gamma {\gamma}' \sigma}
          t_{\gamma {\gamma}'}^{\bf a} d_{{\bf i} \gamma  \sigma}^{\dagger}
         d_{{\bf i}+{\bf a} {\gamma}'  \sigma} -J_h \sum_{\bf i} {\bf s_i} \cdot {\bf S_i} + \nonumber \\
  & &      J_{AF} \sum_{<{\bf ij}>} {\bf S_i} \cdot {\bf S_j}.
\end{eqnarray}

 The Hamiltonian consists of the kinetic energy of the $e_{g}$ electrons with anisotropic 
hopping integrals $t_{\gamma {\gamma}'}^{\bf a}$ ($\gamma$ and ${\gamma}'$ denote the two $e_{g}$ orbitals, 
${\bf i}$ and ${\bf a}$ index the sites and the first neighbours respectively), a Hund coupling which favours 
the alignment of their spins (${\bf s_i}$) with the core-like $t_{2g}$ moments (${\bf S_i}$) and superexchange
interaction between the classical $t_{2g}$ spins. 

The transfer integrals between two orbitals $e_{g(3z^2-r^2)}$ (orbital 1) and  $e_{g(x^2-y^2)}$ (orbital 2) on adjacent Mn ions 
are given by:
\begin{eqnarray}
t_{\gamma {\gamma}'} &=& -t_{0} \left(\matrix{1 & 0 \cr 0 & 0} \right) \qquad \text{\hspace{1.2cm}along $z$} \nonumber \\
t_{\gamma {\gamma}'} &=& -\frac{t_{0}}{4} \left(\matrix{1 & -\sqrt{3} \cr -\sqrt{3} & 3} \right) 
\qquad \text{along $x$} \\
t_{\gamma {\gamma}'} &=& -\frac{t_{0}}{4} \left(\matrix{1 & \sqrt{3} \cr \sqrt{3} & 3} \right) \nonumber
\qquad \text{\hspace{0.6cm}along $y$}, 
\end{eqnarray}
and $t_{0} = \frac{V_{pd\sigma}^2}{e_{p}-e_{d}}$ where $V_{pd\sigma}^2$ is the Slater-Koster parameter and 
$e_{p}$, $e_{d}$ are the energies of the O ${2p}$ and Mn $3d$ states. This parameter takes into account the
hybridisation with O which does not appear explicitly in the model. 

As shown above an electron in the state $e_{g(3z^2-r^2)}$ cannot hop along the $z$ direction whereas its hopping 
integral along $x$ and $y$ is larger than for the $e_{g(x^2-y^2)}$. This fact will be important in the determination
of the occupancy of the two orbitals in the presence of the
surface and/or an interlayer antiferromagnetic coupling. The strong on-site Hund coupling will 
favour the alignment of neighbouring core spins via the itinerant $e_{g}$ electrons. Competing with this 
tendency is the superexchange which acts between core spins on neighbouring sites. This interaction is 
responsible for the observed G-AF phase in the end members AMnO$_{3}$ where the $e_{g}$ electrons are 
absent and thus only the superexchange operates. The superexchange also wins over when there are not enough 
carriers to lower the total energy by a gain in kinetic energy or in case where hopping is suppressed due to 
other factors as is the case in the presence of a surface as we will see below. In order to keep the 
number of parameters to a minimum we did not include the Coulomb on-site repulsion between $e_{g}$ electrons nor
the Jahn-Teller coupling \cite{calderon}. On the other hand we added a shift\cite{calderon} $\Delta$ of the 
on-site energy for the orbitals at the surface in order to take into account the change in energy of the states at the 
surface due chemical shifts and/or strain fields. This may be the case at interfaces with grain boundaries or with insulating
barriers in tunneling devices as explained above. One of the 
major effects of the surface is the occurrence of a charge transfer to or from the bulk region inducing a loss 
of local neutrality and creation of electrostatic dipoles. We treat these interactions in the Hartree 
approximation \cite{calderon} by solving the Poisson equation with $\epsilon=5$. The coupled 
Schr\"odinger-Poisson equations are solved self consistently until the relative change in energy and charge is 
less than 0.05\%.

\section{Results and discussion}
\subsection{Orbital ordering and surface magnetism}
\begin{figure}
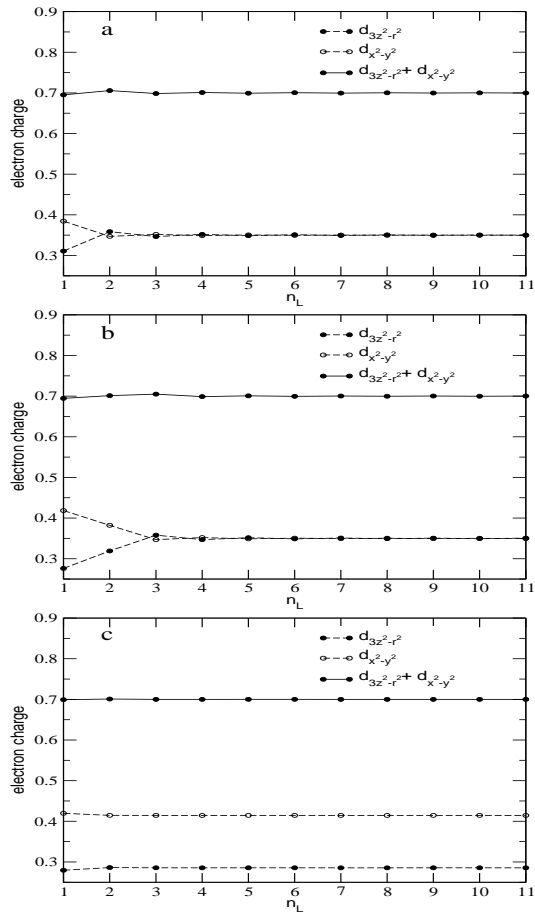

\includegraphics*[width=7cm,height=4cm]{FM_Orbits_delta_0.00.eps}
\includegraphics*[width=7cm,height=4cm]{100-001_Orbits_delta_0.00.eps}
\includegraphics*[width=7cm,height=4cm]{A-AF_Orbits_delta_0.00.eps}
\caption{\label{fig1}
Occupancy of the  $e_{g(3z^2-r^2)}$ and $e_{g(x^2-y^2)}$ and their sum as a function of
the distance from the surface for a 21-layer TB model of R$_{0.7}$A$_{0.3}$MnO$_{3}$. The configurations are
in order FM, FM with the surface layer flipped (DUUD) and A-AF.}
\end{figure}

We will consider three possible magnetic configurations namely ferromagnetic (FM), A-type antiferromagnetic (A-AF)
and a configuration where all the moments on the inner layers are parallel and the surface one flipped (DUUD, where D is for down 
and U is for up) in order to look at the interplay between magnetic ordering and orbital ordering and the effect of the surface. 
In section III.C we evaluate the energy of a reversed layer. We assume an in-plane ferromagnetic ordering so that we have 
one inequivalent atom per plane. We show in  Fig. \ref{fig1} the occupancies of the $e_{g(x^2-y^2)}$ and $e_{g(3z^2-r^2)}$ 
orbitals in these three configurations. There is a noticeable correlation between the type of magnetic and orbital orderings. 
The $e_{g(x^2-y^2)}$ is more populated than the $e_{g(3z^2-r^2)}$ on those planes which are antiparallel to their neighbours. 
Whereas the two are equally occupied when the coupling is FM. The surface layer is an exception however but this is not surprising
having in mind the anisotropy of the transfer integrals defined above. 

The higher occupancy of the $e_{g(x^2-y^2)}$ orbital at the surface is explained by the absence of interlayer 
hopping for the electrons in this state which do not lose kinetic energy in the presence of the surface.
Whereas the $e_{g(3z^2-r^2)}$ electrons are more sensitive to the presence of the surface which limits their
hopping and as a result this orbital is more occupied in the bulk where the levels are broadened than at the 
surface where the level is more localised. As mentioned above the anisotropy of the hopping integrals leads to no
direct electron transfer from $e_{g(x^2-y^2)}$ orbitals between planes. Hence in the current model the local 
density of states (DOS) 
projected on this orbital is independent both of the position in the slab and the magnetic orientation of the
neighbouring planes. On the other hand there is transfer between $e_{g(3z^2-r^2)}$ orbitals along $z$. This means
that the local $e_{g(3z^2-r^2)}$ DOS will be narrowed at the surface because the transfer is only to one plane
instead of two and this is true also if there is AF order. We see this effect clearly  in Fig. \ref{fig1}.
Conversely, the decrease in the kinetic energy of the  $e_{g(3z^2-r^2)}$ electrons at the surface results in the 
weakening of the double exchange and a tendency to an AF coupling between the surface and subsurface layers. The
 higher occupancy of the $e_{g(x^2-y^2)}$ orbital at the surface makes it more likely that it will want to 
localise as we will see below in the SIC calculations.  Inside the bulk-like region of the slab the two orbitals 
are equally likely to be occupied as long as the coupling between layers is FM. The strong on-site Hund's 
coupling will always act to align the $e_{g}$ electron's spin to the $t_{2g}$ one and in order for the system to 
gain from the kinetic energy of the electrons, the core spins ought to be parallel. If this is not the case then 
the hopping is partially suppressed and the superexchange wins over. This suppression occurs between layers and 
as the only orbital which has a finite transfer integral in this direction the $e_{g(3z^2-r^2)}$ will be 
penalised and hence depopulated as can be seen from Fig. \ref{fig1}. In Fig. \ref{fig1} where no shift of the 
surface levels is introduced the $e_{g(3z^2-r^2)}$ is disfavoured near the surface because of the narrowing of
the local DOS. The electrons would rather go to the wider $e_{g(x^2-y^2)}$ DOS. Thus there is an orbital order
induced at the surface and is present in all of the three configurations but is stronger when there is a local
AFM coupling between the surface and subsurface layers as is the case in (b). The effect is even bigger in (c)
where the AF coupling all through the slab causes a narrowing of the $e_{g(3z^2-r^2)}$ and no significant change
to the $e_{g(x^2-y^2)}$ DOS so that we found orbital order throughout the film. The total electron density shown 
by the sum, $d_{3z^2-r^2} + d_{x^2-y^2}$, is almost unchanged at 0.7. This is due to the suppression of charge 
imbalance by including the electrostatic interactions solved for in the Poisson equation. Without this extra electrostatic term 
the electrons in the absence of a shift of the surface levels would transfer to the inner part of the slab
raising the density above the bulk level independently of the slab thickness. We see however from Fig. \ref{fig1}
that we get the bulk properties for three layers away from the surface. 

\begin{figure}
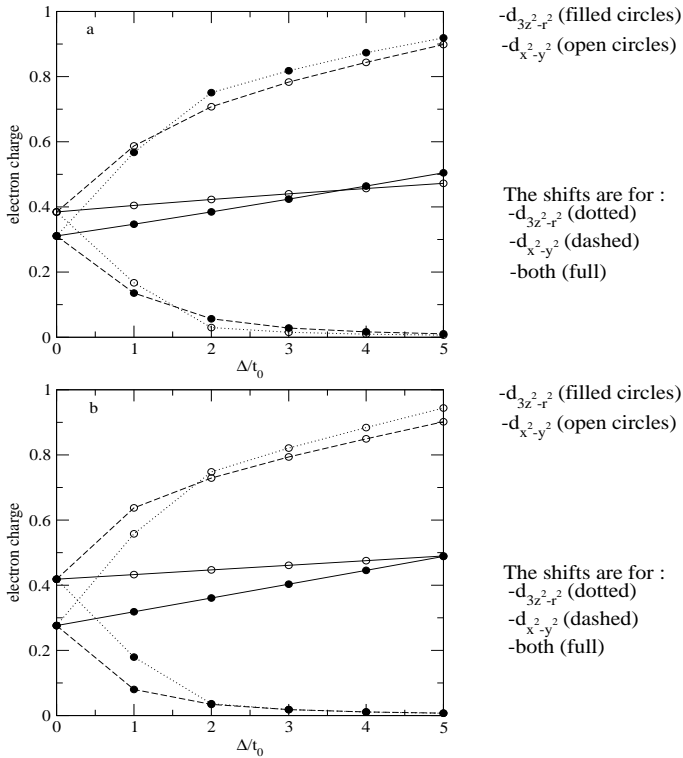

\includegraphics*[width=9cm,height=5cm]{occupancy_vs_Detla_fm.eps}
\includegraphics*[width=9cm,height=5cm]{occupancy_vs_Detla_duu-uud.eps}
\caption{\label{fig2}
Evolution of the occupancies of the $e_{g(3z^2-r^2)}$ and  $e_{g(x^2-y^2)}$ orbitals at the surface
          versus the shift $\Delta$. Two configurations are considered: FM(a) and DUUD (b).}
\end{figure}

\begin{figure}
\includegraphics*[width=7cm,height=4cm]{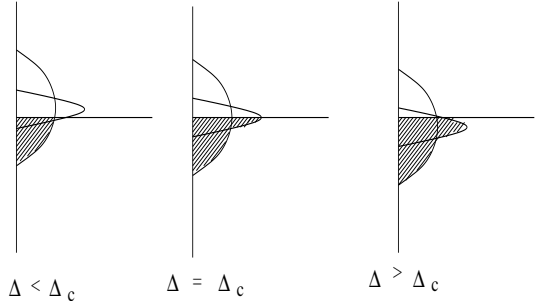}
\caption{\label{fig3}
Schematic representation of the evolution of the occupancy of the  $e_{g(3z^2-r^2)}$ (narrow) and $e_{g(x^2-y^2)}$ (broad) 
orbitals at the surface when shifted by an amount $\Delta$ with respect to their common bulk level. The area below the two curves
is the same. The horizontal line represents the Fermi energy. When the Fermi level is above the centre of the two bands the 
occupation of $e_{g(3z^2-r^2)}$ becomes higher}
\end{figure}

We have studied the effect of introducing the shift $\Delta$ for both orbitals and for only one of them at the 
surface. As explained earlier, this parameter is used to mimic the effect of structural (like lattice mismatch) and 
chemical (like charged surfaces) on the on-site energy level of the orbitals. Here we look at the effect on the orbital ordering and
in Section III.C we will consider the changes in the relative stability of the two solutions FM and DUUD. The orbital occupancies 
are given in Fig.\ref{fig2} as a function of the strength $\Delta$ for the two magnetic solutions corresponding to $(a)$ and $(b)$
of the Fig. \ref{fig1}. We found that when both orbitals are shifted by a small $\Delta$ the occupation of the  $e_{g(x^2-y^2)}$ 
remains higher and is explained from the simple kinetic energy gains argument explained earlier. Increasing $\Delta$ results in a 
crossover to higher occupancy of the $e_{g(3z^2-r^2)}$ orbital but this is explained also from the anisotropy of the transfer 
integrals argument. As is depicted schematically in Fig. \ref{fig3} the LDOS of the $e_{g(x^2-y^2)}$ orbital is broader than that
of $e_{g(3z^2-r^2)}$ so that when the Fermi level is well below the centre of the bands the occupation of the first is higher. 
With increasing $\Delta$ the Fermi level lies at the centre and the two orbitals are equally filled. Increasing further
$\Delta$ will favour the occupation of  $e_{g(3z^2-r^2)}$ which has higher number of states available in a much smaller energy 
window. The crossover occurs for smaller values of $\Delta$ in the FM solution than in the DUUD case, in agreement with our 
earlier finding that the occupation of $e_{g(3z^2-r^2)}$ favours FM coupling. Shifting one orbital only will favour its occupation
in both solutions and for values of $\Delta$ larger than 2$t_{0}$ the other orbital is completely depleted.

\begin{figure}
\includegraphics*[width=7.7cm,height=9cm]{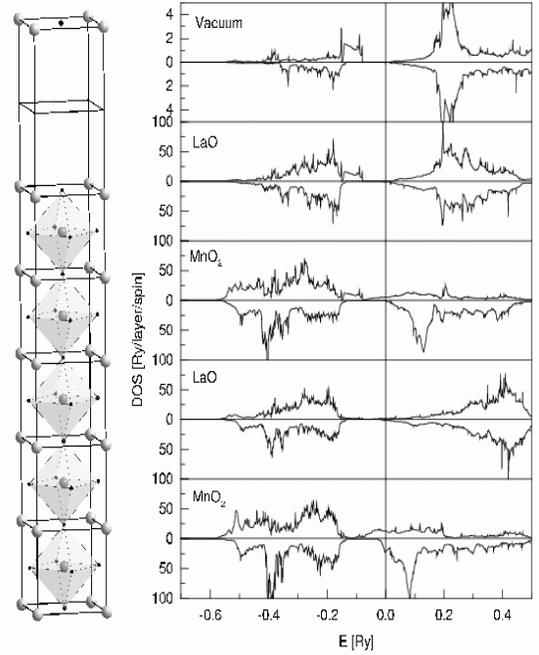}
\caption{\label{fig4}
LDOS, in the rigid band model, for a 5 units supercell (La$_{0.7}$Sr$_{0.3}$MnO$_3$)$_5$.
The Mn in layers 2 and 10 have localised $3t_{2g}+1e_{g(x^2-y^2)}$ electrons and the other Mn have 
localised $3t_{2g}$ electrons only. The left-hand-side picture shows the supercell.}
\end{figure}

\subsection{\textit{Ab initio} study of charge and orbital ordering}

As mentioned above we report also on results using the SIC-LSDA to study the surface of a a representative system,
La$_{0.7}$Sr$_{0.3}$MnO$_{3}$. This method has already been used with success in studying the bulk properties of 
this material \cite{banach}. It allows for a parameter-free total energy minimisation with respect to the 
localised/itinerant state of the Mn $d$ electrons in our case. The number of electrons allowed for band formation
is found by comparing total energies in the two configurations where the electron is itinerant and where it is 
localised. The valency of the Mn ions is then found by subtracting the localised electrons from the total number 
of valence electrons.  It has been applied successfully to systems where there is strong tendency to localisation of 
the valence electrons that could not be accounted for using the conventional LSDA functionals.\cite{strange}

 Here the focus is put on the changes brought about by the surface on the charge and orbital orderings.
The valence of Mn in LSMO was calculated to be tetravalent \cite{banach} in the bulk material.
However we find here that this valence is reduced to trivalent when the Mn is on the subsurface layer.
We report on calculations in which we include a virtual La/Sr atom to account for the mixed
valence. All the layers are either MnO$_{2}$ or La$_{0.7}$Sr$_{0.3}$O. This is a type of rigid band model.
First we studied a supercell consisting of four layers of MnO$_{2}$ with three and four layers of 
La$_{0.7}$Sr$_{0.3}$O. The second system is considered in order to check the effect of the stoichiometry which is
not respected in the system with three La$_{0.7}$Sr$_{0.3}$O layers only. This non-stoichiometric system is symmetric 
and the calculations are much less involved than in the stoichiometric but non-symmetric case. The slabs are
separated by seven and six layers of empty spheres in the symmetric and non-symmetric case respectively. The 
non-symmetric system has two surface terminations, \textit{i.e} MnO$_{2}$ and La$_{0.7}$Sr$_{0.3}$O whereas in the
symmetric case the termination is MnO$_{2}$. 

In the LSDA calculations we found an energy difference of 8.04 mRy/(MnO$_{2}$ layer) between the ground-state FM
configuration and the configuration where the surface moment is antiparallel to the bulk ones (DUUD) in the symmetric
case. For the non-symmetric case the ground state is also FM and the difference in energy with the flipped surface
moment configuration is of 5.59 mRy/(MnO$_{2}$ layer). The surface and subsurface Mn magnetic moments in the ground state
are of 3.26 and 3.16$\mu_B$ in the symmetric system and of 3.28 and 3.12$\mu_B$ in the non-symmetric system. 
The results are indeed in good agreement. We then studied different orbital localisation scenarios. These are the 
localisation of the 3$t_{2g}$ orbitals all through the slab and localisation of an extra $e_{g}$ orbital at the surface, 
which gives three possible scenarios for each of the magnetic configurations. In both systems the ground state is the FM 
phase with the 3$t_{2g}$ orbitals only localised on all the Mn ions. The FM configuration with an extra $e_{g(x^2-y^2)}$
localised at the surface is 6.67 mRy/(MnO$_{2}$ layer) higher in the symmetric case and is of 
6.21 mRy/(MnO$_{2}$ layer) in the non-symmetric case. The surface and subsurface Mn magnetic moments 
are of 3.30 and 3.31$\mu_B$ in the symmetric system and of  3.32 and 3.27$\mu_B$ in the stoichiometric system.

Having confirmed that the stoichiometry has negligible effect on the overall relative 
stability of different orbital and magnetic configurations we applied this method to 
symmetric five-MnO$_2$ layers supercell of LSMO surface which included four unit cells of empty spheres  
(see Fig.~\ref{fig4}). In this case the MnO$_{2}$ layer is at the sub-surface rather than the surface which we studied previously.
The TB model does not include the La$_{0.7}$Sr$_{0.3}$O layers and therefore cannot
differentiate between these cases. However a phenomenological shift $\Delta$ of the on-site energies at the surface could 
capture this. We found that terminations by the La$_{0.7}$Sr$_{0.3}$O layer leads to the localisation of one more 
electron on the Mn atoms in the MnO$_2$ layer under surface. The ground state configuration has localised 
3$t_{2g}+1e_{g(x^2-y^2)}$ electrons under the surface and 3$t_{2g}$ electrons on the 
manganeses in the bulk in agreement with the model calculations when the shift $\Delta$ is applied to the surface
e$_{g(x^2-y^2)}$ orbital (see Fig.~\ref{fig5}). Since the MnO$_{2}$ layer is not the termination, the 
$e_{g(3z^2-r^2)}$ could not be said to be favoured as was the case in earlier calculations \cite{calderon,fillipetti2}. 
There is however an electrostatic interaction with the surface $O^{2-}$ ion which means that the energy level of this 
orbital should increase in comparison to the case when the MnO$_2$ layer is at the surface. Whereas this is not 
true for the e$_{g(x^2-y^2)}$ orbital which is less sensitive to the presence of the extra La$_{0.7}$Sr$_{0.3}$O layer. 
The situation can then be modelled by shifting the on-site e$_{g(x^2-y^2)}$ level downwards which is equivalent to shifting the 
other level upwards. The calculated ground state configuration has the total energy of 37 mRy lower than the system with all 
Mn$^{4+}$ configured manganese. From the LDOS of the first layer we can see small contributions of electronic states 
from the vacuum region. The La$_{0.7}$Sr$_{0.3}$O surface layer becomes insulating with a band gap of about 1 eV. The 
LDOS for the first MnO$_2$ layer with Mn$^{3+}$ valence shows a nearly half-metallic character with a nearby pseudo-gap 
of 1.7 eV. This shows that localisation of the e$_{g(x^2-y^2)}$ electron, forced by the surface, leads to near 
half-metallic properties at the LSMO  surface as was mentioned before \cite{banach}. The change of the symmetry of the 
localised electron to $e_{g(3z^2-r^2)}$ increases the total energy by about 23 mRy. This is a substantial energy, 
62\% of the energy needed for delocalizing this e$_g$ electron.
Additionally, the rotation of $e_{g(3z^2-r^2)}$ orbitals by 90$^\circ$ into the $x-y$ plane 
makes this configuration unfavourable by only 10 mRy in comparison with the ground state
configuration of a localised e$_{g(x^2-y^2)}$ electron.
Localisation of $e_g$ electrons on the other Mn atoms, inside the bulk, 
increases the total energy of the system.
The LDOS for the next LaO layer shows metallic character with 
a small number of electrons at the Fermi level but
each of the MnO$_2$ layers including the Mn$^{4+}$ has a clear metallic character. 
The LDOS for the layers below these are 
similar to the ones for the bulk.
We noted that decreasing the volume of elementary cell increases the 
differences between the energies described above.
However, increasing the lattice parameter by 2\% decreases the 
differences between energies states ten times.
The magnetic order at the surface changes through spin flip processes from the bulk
to a local anti-ferromagnetic arrangement.

\subsection{Energetics}
Depending on the surface/interface termination several scenarios are possible which will allow for a particular
orbital to be favoured. It has been argued that a termination with MnO$_{2}$ plane will favour the occupation
of the e$_{g(3z^2-r^2)}$ orbital \cite{calderon} which has its lobe oriented toward the missing oxygen ion and 
hence has lower electrostatic energy than the e$_{g(x^2-y^2)}$ which still sees the oxygens present in the plane.
A phase diagram in the parameter space has been obtained \cite{calderon} where it is shown clearly
that a large shift $\Delta$ of the e$_{g(3z^2-r^2)}$ with respect to  e$_{g(x^2-y^2)}$ and the bulk levels will
favour an inplane antiferromagnetism at the surface and a canted configuration with respect to the ``bulk''. 
\begin{figure}[t]
\includegraphics*[width=8cm,height=5cm]{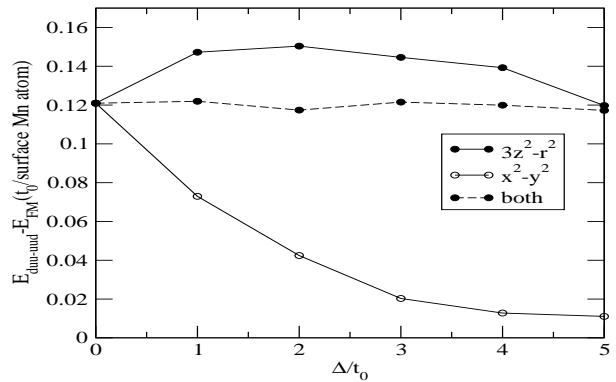}
\caption{\label{fig5} Energy differences between the two configurations where the surface-subsurface coupling
is AFM or FM in units of the hopping parameter $t_{0}$ per surface Mn ion versus the amount $\Delta$ of the shift
added to one or both surface levels. The Hund's coupling parameter is $J_{h} =6t_{0}$.}
\end{figure}
First principles calculations \cite{fillipetti2} on the other hand found that the surface-subsurface exchange 
coupling in La$_{x}$Ca$_{1-x}$MnO$_{3}$ is FM independently of the ``bulk'' magnetic configuration. The authors\cite{fillipetti2} 
argued that this is due to the dangling e$_{g(3z^2-r^2)}$ bond again which favours this ferromagnetism. Whereas in the present 
\textit{ab initio } study we have found that it is rather the  e$_{g(x^2-y^2)}$ that is localised as the MnO$_{2}$ layer is
the sub-surface rather than the terminating layer. 
We present below results concerning the competition between FM and AFM 
surface-subsurface coupling as a function of the symmetry of the localised orbital and the amount $\Delta$ by which the levels 
are shifted. We considered shifts of both levels simultaneously and of one orbital at a time and calculated the band energies 
(kinetic plus Hartree) when the surface-subsurface coupling is either FM or AFM. The superexchange energy difference between 
them is of $2J_{AF}$ per surface Mn ion. The results are shown on Fig.~\ref{fig5} where the energies are given in units
of the hopping integral $t_{0}$ and the Hund's coupling constant $J_{h}$ is taken equal to $6t_{0}$. As can be 
seen from the figure while localising both orbitals does not change the relative energies, localising the 
e$_{g(x^2-y^2)}$ will affect strongly the surface-subsurface ferromagnetism. This is due to the depletion of the 
other orbital on the surface and hence the weakening of the double exchange mechanism mediated by 
e$_{g(3z^2-r^2)}$ electrons hopping between the layers. In the case where both orbitals are shifted it is the 
reduction in the subsurface occupation of the e$_{g(3z^2-r^2)}$ orbital which limits the gain in kinetic energy 
that would result from the higher occupation of the surface e$_{g(3z^2-r^2)}$ orbital. The relative stability of 
the two configurations remains unaltered as a result. Shifting the e$_{g(3z^2-r^2)}$ will enhance the FM coupling
with the bulk as found in the \textit{ab initio} study of the MnO$_{2}$-terminated system. But this is true only if
the e$_{g(x^2-y^2)}$ is not strongly disfavoured as found in the previous model calculations \cite{calderon}.
If the occupation of the  latter orbital is low at the surface in-plane anti-parallel coupling of the spins would result as
a consequence of the weakening of the double exchange at the surface which is mediated mostly by e$_{g(x^2-y^2)}$ electrons.

\section{Conclusion}

We have considered what light our calculations have shed on the ideal 
hole-doped-manganite-insulator interface such that tunneling
magnetoresistance (TMR) is optimal.  
We have studied changes that are induced by the lack of cubic symmetry at the surface as 
well as different chemical environments which favour the formation of localised states, 
through a realistic double exchange model
and first principles calculations. 
In the model calculation we have taken account of the different scenarios by adding a shift $\Delta$ 
to the surface on-site energy of the orbitals. If a surface/tunnel barrier has net positive charge $\Delta$ will be
negative for both orbitals. If this is too large we get localisation which is bad for tunneling. Equally a strong negative
charged termination is also bad because a positive $\Delta$ 
will deplete both orbitals leading to magnetic disorder at the surface. 
If the in-plane lattice constant of the barrier is smaller than that of the manganite crystal 
there will be a strain field which gives a negative $\Delta_{3z^2-r^2}$ 
tending to favour the  e$_{g(3z^2-r^2)}$ orbital. For small values of 
$\Delta_{3z^2-r^2}$ surface ferromagnetism is enhanced. 
However large values of this strain will deplete the e$_{g(x^2-y^2)}$ orbital 
and favour in-plane antiferromagnetism at the surface MnO$_{2}$ layer which is detrimental to TMR. 
On the other hand strains favouring 
e$_{g(x^2-y^2)}$ always suppresses ferromagnetic ordering and hence TMR.
In the absence of any chemical shift of the atomic levels at the surface only a small amount of charge 
is transfered to the bulk because of the resulting electrostatic forces tendency 
to restore local charge neutrality. Shifting both 
levels or one of them however will result in charge transfer to the surface for large values of $\Delta$. 
This leads to the formation of Mn$^{3+}$ at the surface. 
These findings are confirmed by the more accurate \textit{ab initio} SIC-LDA calculation 
on a model system La$_{0.7}$Sr$_{0.3}$MnO$_{3}$. 
In these calculations we found no localisation of the surface orbitals when the 
slab is terminated by  a MnO$_{2}$ layer and the coupling is ferromagnetic. 
We studied the LaSrO-terminated system were the 
subsurface e$_{g(3z^2-r^2)}$ still sees an oxygen ion on the surface and interacts strongly with it. 
This orbital is then disfavoured and we found that the e$_{g(x^2-y^2)}$ is localised at the subsurface layer 
changing the valency of the Mn ion from tetravalent in the bulk to trivalent at the surface. 
The magnetic  coupling then becomes antiferromagnetic as would be expected from the 
correlation between orbital and magnetic ordering. However if the interlayer distance is increased at the surface, which amounts 
to favouring the e$_{g(3z^2-r^2)}$, we found that the energy differences between competing orderings decrease significantly. 
This relates the \textit{ab initio} results to the model ones. In summary we predict that the best TMR will come from tunnel
barriers that are neutral or weakly positive. There should be a minimal surface strain although a small in-plane compressive 
strain favouring the e$_{g(3z^2-r^2)}$ is actually beneficial.

\end{document}